\documentclass[a4paper]{report}
\pdfoutput=1
\usepackage[utf8]{inputenc}
\usepackage[T1]{fontenc}
\usepackage{RJournal}
\usepackage{amsmath,amssymb,array}
\usepackage{booktabs}

\hypersetup{pdfauthor={Federico Marotta},pdftitle={fplyr}}

\begin{document}

\sectionhead{Contributed research article}
\volume{XX}
\volnumber{YY}
\year{20ZZ}
\month{AAAA}

\begin{article}
\title{fplyr: the split-apply-combine strategy for big data in R}
\author{by Federico Marotta}

\maketitle

\abstract{
	We present \pkg{fplyr}, a new package for the R language to deal 
	with big files. It allows users to easily implement the 
	split-apply-combine strategy for files that are too big to fit into 
	the available memory, without relying on data bases nor introducing 
	non-native R classes. A custom function can be applied independently 
	to each group of observations, and the results may be either 
	returned or directly printed to one or more output files.
}

\section{Introduction}

Many fields of science and industry are witnessing an expansion in the 
volume of data collected by their statistical experiments, and the size 
of the files to be analysed is growing accordingly. On the other hand, 
the hardware resources needed to scale up the analyses are not available 
to every practitioner, and sometimes are even beyond the current 
technological reach. In particular, it can happen that the size of a 
file exceeds the total RAM on the machine where the analysis is to be 
performed. In such cases the R \citep{Rlang} programmer has several 
tools at his or her disposal, each with its own advantages and 
limitations. The existing tools can be broadly classified in three 
families: database-backed, file-backed, and MapReduce. In the next few 
paragraphs, these three approaches will be briefly reviewed; the 
purpose, however, is not to compare them, as each is suited to different 
situations.

To the first family, database-backed, belong packages such as 
\CRANpkg{sqldf} \citep{grothendieck2012sqldf} and 
\href{http://rdbi.sourceforge.net/}{\pkg{Rdbi}}, which set up a database 
behind the scenes and perform SQL queries on it without loading the 
whole file into memory. While this approach can leverage the speed and 
efficiency of database routines, it is sometimes necessary to perform on 
the data particular operations that are not entailed by the database 
software. For example, while it is possible to use the \code{groupby} 
SQL keyword to aggregate the results of a query according to the value 
of one field, only a small number of simple `aggregating operations' are 
supported; the exact number and type of operations can differ according 
to the database backend, but in general they are not much more 
sophisticated than summary statistics like \code{AVG}, \code{MAX}, or 
\code{STDEV}.

A filebacking strategy is employed in the \CRANpkg{bigmemory} 
\citep{kane2013bigmemory} and \CRANpkg{ff} \citep{adler2014ff} packages, 
which rely on low-level operating system functions to create a virtual 
map of the file, once again without loading it into memory. These 
packages are highly optimised, but their efficiency comes with 
tradeoffs; for instance, \CRANpkg{bigmemory} is restricted to numerical 
matrices, and with \CRANpkg{ff} only a (wide but) limited set of 
preimplemented operations are available. The \CRANpkg{ffbase} 
\citep{de2014ffbase} considerably enlarges the scope of action of 
\CRANpkg{ff}. One potential disadvantage that remains, however, is that 
\CRANpkg{ff} and related packages implement their own set of classes, 
rather than using the native ones offered by R.

The \CRANpkg{iotools} package \citep{arnold2015iotools} offers both 
an efficient way to read a file from disk and a set of functions to 
parse the file chunk by chunk. These latter functions, such as 
\code{chunk.apply()} and \code{chunk.map()}, are reminescent of the 
MapReduce paradigm, which gives the name to the last family of our 
classification. Here, the large file is read piece by piece in such a 
way that, at any given moment, only a limited number of rows of the file 
are present in the RAM; at the same time, an arbitrary function can be 
applied to each `chunk'. The results of the processing are then combined 
and returned.

In this paper we propose a new approach, implemented in the 
\CRANpkg{fplyr} pacakge, to deal with big files; it aims to be a simple 
and user-friendly solution that integrates well with the existing R 
functions. Nonetheless, as all of the approaches described above, it 
also has a restricted applicability, as it deals only with a particular 
class of big files: those that are amenable to the split-apply-combine 
approach \citep{wickham2011split}. For example, the famous \code{iris} 
data set \citep{anderson1936species,fisher1936use} would be a good 
candidate because the `Species' field defines a partition of the data 
into independent groups of observations, so that a particular function 
can be applied separately to each species. Another example is that of 
gene expression files, which are often organised as matrices where each 
row refers to a gene, and each column reports the expression of that 
gene in a different individual. In this case, one may apply a function 
separately to each gene. More specifically, the files on which 
\CRANpkg{fplyr} operates should be formatted in such a way that 
consecutive rows contain all the measurements relative to the same 
subject, and the first field contains the subject IDs. We refer to each 
group of consecutive rows pertaining to the same subject as a 
\dfn{block}.

Much as \code{apply()} applies a function to each row or column of a 
matrix, the functions in \CRANpkg{fplyr} apply custom functions to each 
block of data, independently of all the other blocks. Thus, at its core, 
this package enables one to mimick the behaviour of the \code{by()} 
function, without requiring that the whole file be loaded into memory. 
Indeed, the file being read block by block, \CRANpkg{fplyr}'s functions 
run with an $O(1)$ space complexity.

As an illustration of the possible usage of the package, suppose that 
the path to a big file to be processed is stored as a character string 
in the variable \code{f}. Then, the following code computes and returns 
the \code{summary()} of each block:

\begin{example}
flply(f, summary)
\end{example}

\section{Comparison with existing packages}

The approach we presented may appear similar to \CRANpkg{iotools}' 
chunk-wise processing, but it differs from it in several respects. Most 
importantly, one of \CRANpkg{iotools}' chunks can contain measurements 
about many subjects, while an \CRANpkg{fplyr}'s block contains each and 
every measurement about one subject only. In other words, one of 
\CRANpkg{iotools}' chunks may contain many of \CRANpkg{fplyr}'s blocks. 
Moreover, in \CRANpkg{fplyr}, each block is treated as independent of 
all the others; the aim is to obtain a list of values, one for each 
block, whereas with \CRANpkg{iotools} the typical aim is to obtain a 
single value for the whole file.

Similar differences distinguish \CRANpkg{fplyr} by \CRANpkg{ffbase}'s 
way of performing operations by chunks, a function called 
\code{ffdfdply()}. As with \CRANpkg{iotools}, here a chunk can contain 
measurements from many subjects, and the task of further separating them 
is left to the user. Moreover, \code{ffdfdply()} can only return an 
\code{"ffdf"}, the equivalent of a \code{"data.frame"} for \CRANpkg{ff}, 
whereas with \CRANpkg{fplyr} it is possible to return any R object, or 
even to directly write something to an output file block by block, 
during the processing.

As previously stated, aggregate operations with database-backed packages 
are limited to simple operations. If the operation to be applied is more 
complex, it is still possible to use \CRANpkg{sqldf}, but the analysis 
must be performed in at least two steps: first the relevant group of 
observations must be selected, and only then can the custom function be 
applied. In order to replicate the behaviour of the \code{groupby} 
keyword with arbitrary aggregating functions, a loop should be manually 
set up where at each iteration a different group of observations is 
retrieved and analysed; furthermore, all the possible values of the 
`groupby' field must be known in advance. \CRANpkg{fplyr} offers an 
effortless way of doing the same thing.

\section{Implementation}

\CRANpkg{fplyr} is mainly built upon two other R packages: 
\CRANpkg{iotools} and \CRANpkg{data.table}, the former being used to 
read the file chunk by chunk, the latter providing some efficient ways 
to split the chunk into its constituent blocks, applying the 
user-specified function to each, combining the results, and (when 
applicable) writing back the result to the disk \citep{dowle2019}. From 
this point onwards, for the sake of distinguishing the user-specified 
function to be applied to each block from other functions, we shall 
refer to it as \code{FUN}. \CRANpkg{fplyr}'s algorithm is, in essence, 
as follows: one chunk is read with the help of the \CRANpkg{iotools} 
package, then it is split into its constituent blocks using the 
\code{by()} function, and \code{FUN} is applied to each. Once the whole 
file has been processed, it is returned a list where each element 
corresponds to a block. This algorithm, with additional technicalities, 
is implemented in the \code{flply()} function. However, if the output of 
\code{FUN} is a \code{"data.frame"}, the \CRANpkg{data.table} package 
allows us to replace the base R \code{by()} function with a faster 
alternative; this second algorithm is implemented in the \code{ftply()} 
function (see \nameref{sec:names} for an interpretation of the names of 
the functions in this package). Since the user knows \emph{a priori} the 
type of output that \code{FUN} returns, he or she can choose which 
function to use accordingly. Note, however, that the second approach is 
implemented only as a shortcut, because the same result could be 
achieved using \code{by()} followed by \code{rbind()}, albeit it would 
be achieved much more slowly. In the following paragraphs we shall 
discuss several aspects of the implementation.

\begin{figure}[htbp]
  \centering
  \includegraphics[width=.67\textwidth]{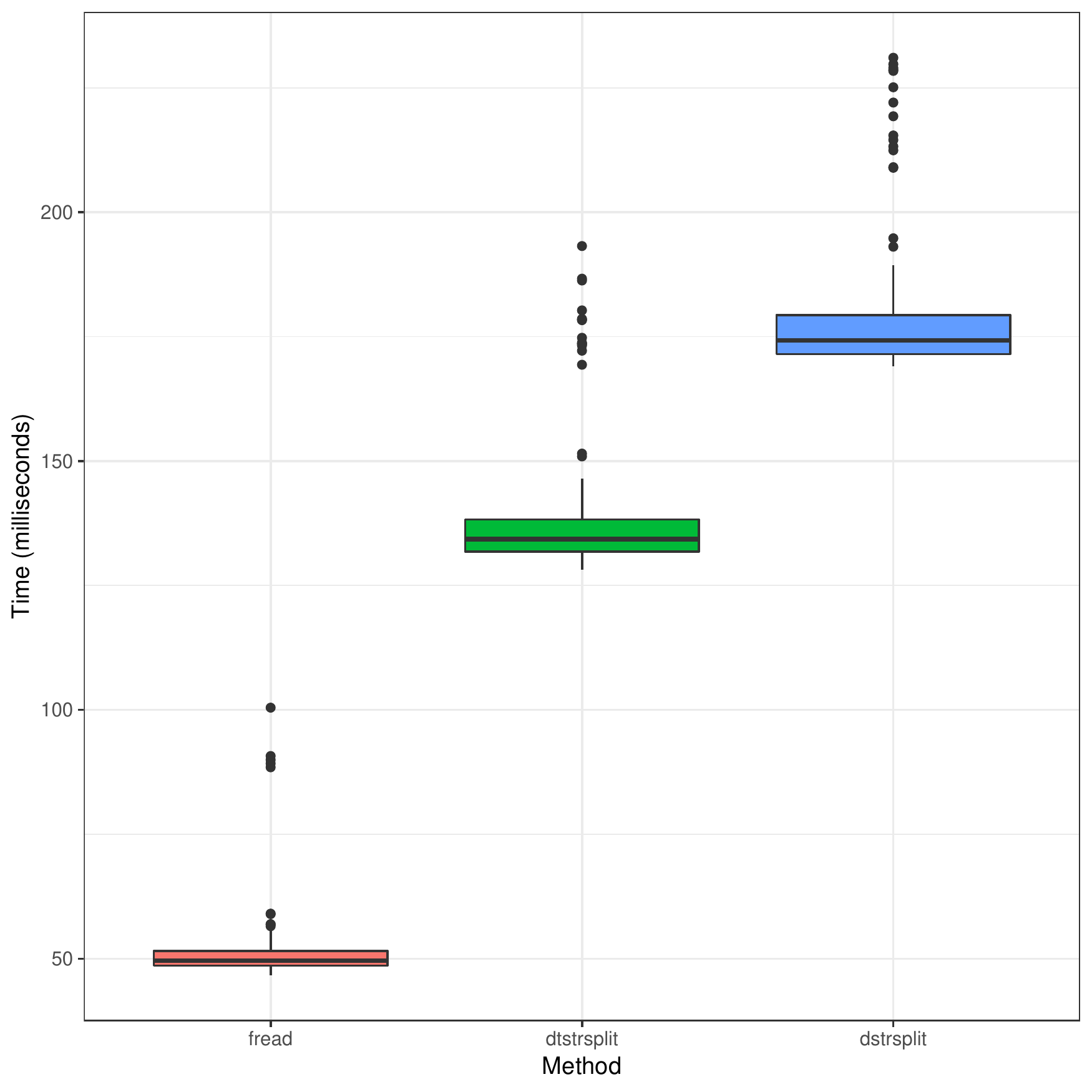}
  \caption{Benchmark of the two formatters: \CRANpkg{fplyr}'s, called 
  \code{dtstrsplit}, and \CRANpkg{iotools}', called \code{dstrsplit}. 
  The two were used to read the same file (253316 lines, 9.2 MB) for 200 
  times without performing any other operation. As a reference, the time 
  taken by \CRANpkg{data.table}'s \code{fread()} is also reported. The 
  benchmark was performed on an Acer Aspire 5750G, 2.4GHz Intel Core i5 
  with 4 cores, 4GB DDR3 RAM. The code to reproduce the benchmarks is 
  available on the GitHub repository of the package at 
  \url{https://github.com/fmarotta/fplyr}.}
  \label{fig:dtstrsplit}
\end{figure}

One of the reasons why \CRANpkg{iotools} is faster than base R functions 
in reading a file from disk is that it reads the file in bulk as a 
\code{"raw"} vector and only afterwards does it convert the data into a 
suitable structure, like a \code{"data.frame"} or a \code{"matrix"}. In 
fact, for instance, the \code{mstrsplit()} function takes a \code{"raw"} 
or \code{"character"} vector and arranges it into a matrix, and 
\code{dstrsplit()} returns a \code{"data.frame"} instead. Functions like 
\code{mstrsplit()} and \code{dstrsplit()} are called \dfn{formatters} in 
\CRANpkg{iotools} jargon \citep{arnold2015iotools}. In \CRANpkg{fplyr}, 
we defined a new formatter which takes a \code{"raw"} vector and returns 
a \code{"data.table"}. In particular, it first converts the \code{"raw"} 
vector to a \code{"character"} one, and then feeds the 
\code{"character"} vector to \CRANpkg{data.table}'s \code{fread()}. 
(Indeed, not only can \code{fread()} take the path to a file as input, 
but it can also handle character strings directly and cast them into 
\code{"data.table"}s.) By relying on it for the low-level reading of the 
files, \CRANpkg{fplyr} inherits some of \CRANpkg{iotools}' features, 
such as the ability to read compressed files. Incidentally, benchmarks 
show that our \CRANpkg{data.table}-based formatter, \code{dtstrsplit()}, 
is even faster than \CRANpkg{iotools}' builtin formatter (Figure 
\ref{fig:dtstrsplit}).

Once a chunk has been read and formatted to a \code{"data.table"}, it 
must be split into its blocks and \code{FUN} must be applied to each of 
them. According to the type of \code{FUN}'s output, as we mentioned, two 
different strategies are employed. If \code{FUN} returns a 
\code{"data.table"}, then we use the following construct:

\begin{example}
output <- chunk[, FUN(.SD, .BY, ...), by = blocks]
\end{example}

In this case, \code{FUN} must take at least two arguments: at evaluation 
time, when \code{FUN} is called, a \code{"data.table"} containing one 
block (except the first field) is passed to \code{FUN} as the first 
argument, and a \code{"character"} vector with the name of the block is 
passed as the second argument. Additional arguments can be specified by 
the user, much as in the \code{apply()} family. It is up to the user to 
write \code{FUN} satisfying these specifications.

On the other hand, if \code{FUN} returns something different than a 
\code{"data.table"}, then \code{by()} is used and the result is returned 
as a list where each element corresponds to a block. In this case, 
\code{FUN} must take at least one argument: at evaluation time, the 
whole block, including the first field, is passed to \code{FUN}. Once 
again, it is up to the user to write or find a function that acts on 
each block and satisfies this specification, and additional arguments 
can be passed by the user like in the previous case.

Due to the mutual independence of the blocks, the processing of a file 
can be easily parallelised\footnote{The parallelisation is only possible 
on *nix operating systems.}; indeed, all the functions in 
\CRANpkg{fplyr} support the \code{parallel} argument to specify the 
number of workers to be initialised. In particular, we adopted 
\CRANpkg{iotools}' \dfn{pipeline parallelism} 
\citep{arnold2015iotools}, where the master process reads from and 
writes to the disk sequentially, but each chunk is then relegated to the 
workers, which process it in parallel as described previously; while the 
workers are busy, the next chunk is pre-allocated by the master. The 
sequential reading and writing avoids conflicts---for instance if 
multiple processes try to write to the same file at the same time---but 
can result in bottlenecks if the processing of each chunk takes an 
amount of time that differs too much from the time needed for the 
reading and writing. Furthermore, some workers can become idle if the 
amount of time required to process different chunks is different.

For convenience, we also implemented two additional functions which may 
be useful when even the result of the processing of the file would not 
fit into the RAM. In these cases, it is not possible to simply return 
everything at the end. If the output of \code{FUN} is a 
\code{"data.table"}, one solution could be to print the resulting data 
for each block to an output file as soon as it is ready, proceeding to
append the results of all the blocks to the same file. This strategy is 
implemented in the \code{ffply()} function. The last function, 
\code{fmply()}, combines the behaviours of \code{ffply()} and 
\code{flply()}: for each block, it is possible to write to one or more 
files and, at the same time, to return an arbitrary R object.

\section{Examples}

To illustrate the usefulness of the package, we shall now discuss some 
examples. Throughout this section, we assume that the path to the file 
of interest is stored in a variable called \code{fin}, while the path to 
an output file is stored inside \code{fout}.

Tabular data with multiple measurements can be represented in two ways: 
the long and the wide format \citep{wickham2007reshaping}. These two 
representations are equivalent in that data can always be converted from 
one to the other, but when the data are in long format, they are `tidy' 
and easier to analyse \citep{wickham2014tidy}, and some R functions 
require their input to be in this format. However, if some big file is 
only available in wide format, reshaping it could become a problem due 
to the scarcity of RAM. Indeed, even if the wide-formatted file fits 
into the memory, the long one may not. With \CRANpkg{fplyr} the 
reshaping can be performed block by block:

\begin{example}
ffply(fin, fout, function(d, by) melt(d, measure.vars = names(d))
\end{example}

Here the \code{by} argument, which contains the subject ID, was ignored, 
but in principle it can be used inside an \code{if} condition to select 
only some of the blocks. The \code{d} argument contains one whole block 
\emph{without the first column}, and it is reshaped using 
\CRANpkg{data.table}'s \code{melt()} function. The first column, 
containing the IDs, will be automatically added after \code{FUN} has 
been applied.

Sometimes a group of observations must be analysed in many ways, 
producing different outputs. For instance, when performing a linear 
regression by blocks, it may be convenient to print the coefficients to 
an output file, while at the same time returning the full \code{"lm"} 
object for future in-depth inspection. The \code{fmply()} function takes 
as arguments the path to the input file, a vector of paths to the 
(possibly many) output files, and a function that returns a list of 
objects. If there are, say, \code{d} output files, the first \code{d} 
elements of the list must be \code{"data.table"}s (or 
\code{"data.frame"}s) and are printed to the corresponding output files; 
optionally, \code{FUN} can return \code{d+1} elements, in which case the 
last one is returned by \code{fmply()}. In this example, the 
coefficients are written to \code{fout}, and the \code{"lm"} objects 
themselves are returned at the end. In particular, \code{l} will be a 
list where each element is an \code{"lm"} object corresponding to one 
block.

\begin{example}
l <- fmply(fin, fout, function(d) {
  lm.fit <- lm(Y ~ ., data = d[, -1])
  # Add the name of the block as the first field
  lm.coef <- as.data.table(cbind(d[1, 1], t(coef(lm.fit))))
  # The coefficients will be printed, the fitted object will be returned
  return(list(lm.coef, lm.fit))
})
\end{example}

Additional examples can be found in the package vignette.

\section{The names of the functions}
\label{sec:names}

Despite the names of this package is reminescent of other packages 
belonging to the tidyverse \citep{wickham2019welcome}, \CRANpkg{fplyr} 
bears no relation with them. We did, however, follow the same 
conventions of one of Hadley Wickham's packages for naming the functions 
\citep{wickham2011split}. All the names consist of two letters followed 
by `ply': the first letter represent the type of input, whereas the 
second letter characterises the type of output, and the final `ply' 
clinches the relation with the existing `apply' family of functions. The 
first letter is usually `f', because the input is the path to a file. 
The second letter is `l' if the output is a list, as in \code{flply()}, 
it is `t' if the output is a \code{"data.table"}, `f' if it is another 
file, and `m' if it can be multiple things.

\section{Discussion}

We believe that \CRANpkg{fplyr} fills a gap in the landscape of the 
existing tools to process big files in R. It adresses a problem that in 
principle could be solved by other packages as well, but only with 
workarounds. Furthermore, our implementation combines the strengths of 
two other packages, \CRANpkg{iotools} and \CRANpkg{data.table}, and is 
therefore reasonably efficient. A variety of features, such as the 
transparent parallelisation, the ability to read compressed files, and 
the possibility to specify the maximum number of blocks, make it also 
user-friendly.

There are, however, also some limitations. First and foremost, the file 
on which \CRANpkg{fplyr} operates must contain observations which can be 
assigned to several independent `subjects', and the subject IDs can only 
be in the first field, the reason being that this is also how 
\CRANpkg{iotools} works. Although it is possible to pre-process the file 
with *nix command-line tools such as awk and sort to ensure that the IDs 
are in the first column and that the rows referring to the same subject 
are adjacent, possible future work could extend the package in order to 
support a custom field for the subject IDs. A related extension could 
allow the blocks to be defined by the combined values of two or more 
columns.

Another possible weakness is the pipeline parallelism algorithm, which, 
besides not being available for Windows-based operating systems, can 
cause bottlenecks if the times required for the operation on the block 
on the one hand, and for reading the block on the other, are on very 
different scales. Nevertheless, this algorithm has been proven to be 
more efficient in a variety of common situations 
\citep{arnold2015iotools}.

In summary, the main strength of the package is that it simplifies the 
task of splitting, applying and combining for files too big to fit in 
the available memory.

\section{URLs}

\begin{itemize}
	\item \href{https://github.com/fmarotta/fplyr}{GitHub repository}
	\item \href{https://cran.r-project.org/package=fplyr}{CRAN page of 
	the package}
	\item \href{https://cran.r-project.org/web/packages/fplyr/vignettes/fplyr.html}{Package 
	vignette}
\end{itemize}

\section{Aknowledgements}

The author would like to thank Prof. Paolo Provero (Università degli 
Studi di Torino) for his insightful comments and suggestions.

\bibliography{fplyr}

\address{Federico Marotta\\
  Università degli Studi di Torino\\
  I-10126, Torino (TO), Italy\\
  ORCiD: 0000-0002-0174-3901\\
  \email{federico.marotta@edu.unito.it}}

\end{article}

\end{document}